\documentclass[prd,nofootinbib,floats,aps,twocolumn,tightenlines,superscriptaddress]
{revtex4-2}
\usepackage{natbib}
\usepackage{graphicx}
\usepackage{mathtools}
\usepackage[normalem]{ulem}
\usepackage{hyperref}
\hypersetup{
	colorlinks=true,
	linkcolor=blue,
	filecolor=magenta,      
	urlcolor=cyan,
}
\usepackage{commath}
\usepackage{bm}
\usepackage{amsfonts,amsmath,amssymb,amsthm}
\usepackage[usenames,dvipsnames]{xcolor}
\usepackage{dsfont}
\usepackage{enumitem}
\usepackage{color}
\usepackage{tikz}
\usepackage{physics}
\usepackage{verbatim}
\newcommand{\proj}[2]{| {#1} \rangle\!\langle {#2} |}

\newcommand{\spt}{\mathcal{M}}

\renewcommand*\d[2][]{%
	\mathrm{d}%
	\ifx\relax#1\relax\else
	\rule{-0.02em}{1.5ex}^{#1}\rule{0.08em}{0ex}\!
	\fi
	#2\,
}

\renewcommand{\dd}{\mathrm{d}}
\newcommand{\rhoh}{\hat{\rho}}
\newcommand{\msf}{\mathsf}
\newcommand{\tsc}{\textsc}

\newcommand{\hilb}{\mathcal{H}}

\newcommand{\tc}{\textsc}

\makeatletter 

\renewcommand\onecolumngrid{
	\do@columngrid{one}{\@ne}%
	\def\set@footnotewidth{\onecolumngrid}
	\def\footnoterule{\kern-6pt\hrule width 1.5in\kern6pt}%
}

\renewcommand\twocolumngrid{
	\def\footnoterule{
		\dimen@\skip\footins\divide\dimen@\thr@@
		\kern-\dimen@\hrule width.5in\kern\dimen@}
	\do@columngrid{mlt}{\tw@}
}%
\makeatother

\newcommand\restr[2]{{
		\left.\kern-\nulldelimiterspace 
		#1 
		\vphantom{\normal|} 
		\right|_{#2} 
}}

\newcommand{\R}[1]{\mathbb{R}^{#1}}

\definecolor{ForestGreen}{RGB}{35,120,35}

\newcommand{\tgg}[1]{{\color{ForestGreen}#1}}

\newtheorem*{claim*}{Claim}

\begin{document}
	
\title{State updates and useful qubits in relativistic quantum information}
	
\author{Jos\'{e} Polo-G\'{o}mez}
\email{jose.polo.gomez@mpq.mpg.de}
\affiliation{Max-Planck-Institut f\"ur Quantenoptik, Hans-Kopfermann-Str. 1, D-85748 Garching, Germany}
\affiliation{Department of Applied Mathematics, University of Waterloo, Waterloo, Ontario, N2L 3G1, Canada}
\affiliation{Institute for Quantum Computing, University of Waterloo, Waterloo, Ontario, N2L 3G1, Canada}
	
\author{T. Rick Perche}
\email{rick.perche@su.se}
\affiliation{Nordita,
KTH Royal Institute of Technology and Stockholm University,
Hannes Alfvéns väg 12, 23, SE-106 91 Stockholm, Sweden}

\author{Eduardo Mart\'{i}n-Mart\'{i}nez}
\email{emartinmartinez@uwaterloo.ca}
\affiliation{Department of Applied Mathematics, University of Waterloo, Waterloo, Ontario, N2L 3G1, Canada}
\affiliation{Institute for Quantum Computing, University of Waterloo, Waterloo, Ontario, N2L 3G1, Canada}
\affiliation{Perimeter Institute for Theoretical Physics, Waterloo, Ontario, N2L 2Y5, Canada}

\begin{abstract}
We address the longstanding challenge of consistently updating quantum states after selective measurements in a relativistic spacetime. Standard updates along the future lightcones preserve causality but break correlations between causally disconnected parties, whereas updates along the past lightcone either imply retrocausality or do not respect the causal propagation of information. We introduce a minimal extension of multipartite states to encode subsystem-specific contextual information. This ``polyperspective'' formalism ensures causally consistent covariant state updates, preserves multipartite correlations, and respects conservation laws. 
\end{abstract}

\maketitle
	
\noindent\textit{\textbf{Introduction.---}} Because not all measurements are destructive, a complete quantum measurement theory requires an \textit{update rule} that specifies the state of a system after a measurement. In non-relativistic quantum mechanics, this is provided by the projection postulate (L\"uders' rule~\cite{Luders1951}). However, relativity constrains a)~what measurements can be performed~\cite{Jubb2022,Albertini2023}, b)~how to model them~\cite{FewsterVerch,Fewster2020covariant,AlexSmith2019Thesis,MeasurementTheory,MeasurementsEH,DanIreneML,Oeckl2013,Oeckl2019,Oeckl2025Spectral,OeacklZampeli2025}, and c)~how to update states after an outcome has been obtained~\cite{Sorkin1993,AharonovAlbert1981}. 
There is no consensus on how to implement updates in relativistic settings~\cite{Hellwig1970formal,AharonovAlbert1980}, even in the simplest case of a pair of qubits in causally-disjoint spacetime regions. 


For instance, consider an EPR pair shared between two parties. If an ideal measurement is performed on one subsystem, L\"uders' rule prescribes updates on both parties. However, when the measurement is embedded in spacetime, it is unclear where to update the state. Should one project the state on a Cauchy surface? Should one only update in the causal future of the measurement? Maybe something else? This question becomes particularly loaded when both parties measure their state while spacelike separated, hence in a scenario where ``who measures first'' is not well defined. The problems that may arise with causality in this context are typically brushed off by noting that the non-local state update does not allow for signalling, since that requires classical communication, which  is assumed to be subluminal. However, while this guarantees that no causality violations will be observed, it says nothing about the physical mechanism through which the state is updated.

In~\cite{Hellwig1970formal}, $\text{Hellwig and Kraus}$ argued that in relativistic quantum theories ideal measurements should alter the state everywhere outside their causal past. However, $\text{Aharonov and Albert}$ later observed that Hellwig and Kraus's prescription failed to respect the conservation of total charge, and concluded that ``no relativistically satisfactory version of the collapse postulate can be found''~\cite{AharonovAlbert1980}.

The inconsistencies raised in~\cite{AharonovAlbert1980} are not problematic under an epistemic interpretation of the quantum state: as $\text{Fewster and Verch}$ put it, ``there is no physical change in the state [...] but rather a shift in which state is appropriate for making predictions, given the information obtained from a measurement''~\cite{Fewster2025MeasurementinQFTEncyclopedia}. Even under this interpretation, relativistic quantum mechanics demands a \textit{relativistic} update rule. That is, \textit{selective} updates must still capture the causal propagation of both the perturbations induced by measurements and the information extracted from them. 

Here, we argue that previously proposed candidates for relativistic state updates are unsatisfactory: a)~an update along the causal future of the measurement cannot account for multi-party correlations and b)~an update on the causal past of the measurement is equivalent to a retroactive update, which should be avoided. 

Then, we propose a covariant framework for state updates that is compatible with relativity. Specifically, we minimally extend the notion of state to include the partial state of information of agents handling different subsystems. This framework allows for the implementation of updates that naturally incorporate how information propagates in spacetime while keeping track of correlations between subsystems.

Finally, we discuss how the framework connects with observer-dependent updates (like those proposed for QFT~\cite{MeasurementTheory}), and analyze how it addresses the problem of conservation of total charge identified in~\cite{AharonovAlbert1980}.
	
\noindent\textit{\textbf{Updating a Bell pair.---}} Let $(\mathcal{M},\msf{g})$ be a (1+$d$)-dimensional globally hyperbolic spacetime, and let $\text{A and B}$ be two qubits following  timelike trajectories $\msf{x}_\textsc{a}(\tau_\textsc{a})$ and $\msf{x}_\textsc{b}(\tau_\textsc{b})$, parametrized by their proper times $\tau_\textsc{a}$ and $\tau_\textsc{b}$, respectively. We assume that they start in the entangled state
\begin{equation}
    \ket{\Psi^+} = \frac{1}{\sqrt{2}}(\ket{0_\tsc{a}}\ket{1_\tsc{b}}+\ket{1_\tsc{a}}\ket{0_\tsc{b}}) \in \hilb_\tsc{a} \otimes \hilb_\tsc{b},\label{eq:EPR}
\end{equation} 
where $\ket{0_j}$ and $\ket{1_j}$ are the eigenstates of the Pauli operator $\hat\sigma_{z,j}$ ($j\in\{\tsc{A},\tsc{B}\}$) associated with the eigenvalues $+1$ and $-1$, respectively. 

Now, assume that at $\tau_\tsc{a}=\tau_\tsc{a}^*$, an ideal measurement of $\hat\sigma_z$ is performed on A, yielding $+1$. In non-relativistic quantum mechanics, there is an absolute time parametrizing the evolution of all systems at once, and the post-measurement state becomes $\ket{0_\tsc{a}}\ket{1_\tsc{b}}$, with partial states $\ket{0_\tsc{a}}$ and $\ket{1_\tsc{b}}$. However, in relativistic setups, there is no absolute time. While we still need to update the joint state following L\"uders' rule to guarantee the compatibility between sequential measurements, we must choose the hypersurface along which this update shall be performed. 

A natural analogue to using an absolute time parameter is to update the state along the hypersurface containing the measurement event $\msf{x}_\tsc{a}(\tau_\tsc{a}^*)$, relative to some spacetime foliation. This prescription is non-covariant, since it depends on a specific foliation. The only hypersurfaces along which the update can be performed covariantly are the future and past lightcones of the measurement event\footnote{For any given point $\msf x \in \mathcal M$, the only set of curves passing through $\msf x$ that generate a $d$-dimensional manifold—without introducing additional structure—are the null geodesics. Depending on whether the tangent vector at $\msf x$ is future-directed or past-directed, these null geodesics generate the future light cone, $\partial \mathcal J^+(\msf x)$, or the past light cone, $\partial \mathcal J^-(\msf x)$, respectively. Any other hypersurface requires invoking a preferred foliation or additional structure, such as Killing vector fields, whose existence is not generally guaranteed in a globally hyperbolic and simply connected $(\mathcal M, \msf g)$~\cite{Wald1984GR}.}.

\begin{figure}
\begin{center}
\includegraphics[scale=1.4]{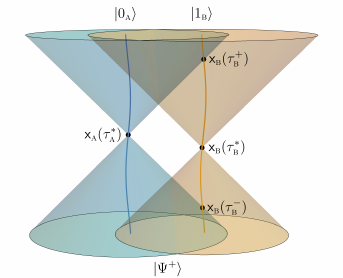}
\caption{Schematic representation of the Bell pair example.}
\label{Fig: measuring a Bell pair in spacetime}
\end{center}
\end{figure}

To assess whether either of these options is adequate, let $\tau_\tsc{b}^- \text{ and  }\tau_\tsc{b}^+$ be the proper times at which B's worldline intersects the past and future lightcones of A’s measurement, respectively\footnote{If B remains outside the future of A's measurement, the discussion still applies setting $\tau_\tsc{b}^+ = +\infty$. It is also worth remarking that for some spacetimes $\tau_\tsc{b}^+$ and $\tau_\tsc{b}^-$ might not be unique. In those cases, we choose $\tau_\tsc{b}^+$ and $\tau_\tsc{b}^-$ to be, respectively, the earliest and latest proper times where B's wordline intersects the measurement event's future and past lightcones.} (see Fig.~\ref{Fig: measuring a Bell pair in spacetime}). If one chooses to update along the past lightcone, the partial state of B changes from
$\hat\rho_\tsc{b} = \Tr_\tsc{a}\proj{\Psi^+}{\Psi^+} = \mathds{1}_\tsc{b}/2$ 
for $\tau_\tsc{b} < \tau_\tsc{b}^-$, to $\ket{1_\tsc{b}}$ for $\tau_\tsc{b} > \tau_\tsc{b}^-$. If we choose the future lightcone instead, the change would happen at $\tau_\tsc{b} = \tau_\tsc{b}^+$. 

In~\cite{Pranzini2023}, it was argued that updating along the past lightcone is the only viable choice, because it is the only one that correctly yields B's statistics conditioned to A's outcome. Namely, because the initial state of A and B is $\ket{\Psi^+}$, if the outcome of measuring $\hat\sigma_{z,\tsc{a}}$ results in +1, a measurement of $\hat\sigma_{z,\tsc{b}}$ can only possibly yield $-1$. According to~\cite{Pranzini2023}, this should be the prediction yielded by B's partial state, even if B has no knowledge of A's measurement outcome---and in particular, even if the measurement of $\hat\sigma_{z,\tsc{b}}$ is performed before $\tau_\tsc{b}^+$---which requires the update to happen at $\tau_\tsc{b}=\tau_\tsc{b}^-$. The problem, however, is that this argument can be carried on to justify a retrocausal update: Assuming that the partial state of B is $\ket{1_\tsc{b}}$ for $\tau_\tsc{b} > \tau_\tsc{b}^-$, one can draw the lightcone of $\msf{x}_\tsc{b}(\tau_\tsc{b}^*)$ for some $\tau_\tsc{b}^* \in (\tau_\tsc{b}^-,\tau_\tsc{b}^+)$ (see Fig.~\ref{Fig: measuring a Bell pair in spacetime}) and use the initial entanglement of A and B to conclude that the partial state of A must be $\ket{0_\tsc{a}}$ \textit{even before the measurement on A is performed}\footnote{This does not imply a contradiction: once we condition our predictions on obtaining a +1 outcome for $\hat\sigma_{z,\tsc{a}}$ at $\tau_\tsc{a}=\tau_\tsc{a}^*$, updating A’s state to $\ket{0_\tsc{a}}$ even for $\tau_\tsc{a} < \tau_\tsc{a}^*$ ensures the correct statistics for earlier measurements, provided the measured observables commute with $\hat\sigma_{z,\textsc{a}}$, which all measurements on B satisfy. While retroactive updates will not lead to inconsistencies under these conditions, they imply accepting that either measurement results are predetermined, or that the whole spacetime is post-selected to be compatible with measurement outcomes---in which case, the state update loses its meaning as a physical process where the system goes from a pre-measurement to a post-measurement state.}.

In contrast, an update along the future lightcone respects the causal structure by fiat and accounts for where the information about the outcome of the measurement is available. However, it fails to account for the correlations between subsystems: Consider, for instance, a second measurement of $\hat\sigma_{z,\tsc{a}}$ performed at some $\tau_\tsc{a}^{**} > \tau_\tsc{a}^{*}$, and a measurement of $\hat\sigma_{z,\tsc{b}}$ performed on B while spacelike separated from $\msf x_\tsc{a}(\tau_\tsc{a}^*)$, at some $\tau_\tsc{b}^* \in (\tau_\tsc{b}^-,\tau_\tsc{b}^+)$. Then, an update along the future lightcone of the first measurement on A predicts the following expectation values: 
\begin{align}\label{Eq: marginal expectation values}
    \langle\hat\sigma_{z,\tsc{a}}\rangle = 1 \:\:\text{ at }\:\: \msf x_\tsc{a}(\tau_\tsc{a}^{**}), \quad
    \langle\hat\sigma_{z,\tsc{b}}\rangle = 0 \:\:\text{ at }\:\: \msf x_\tsc{b}(\tau_\tsc{b}^*),
\end{align}
and yet, because A and B are initially correlated, it must hold that
\begin{equation}\label{Eq: correlations}
    \langle\hat\sigma_{z,\tsc{a}} \otimes \hat\sigma_{z,\tsc{b}} \rangle = -1.
\end{equation}
No joint density operator \mbox{$\rhoh_\tsc{ab} \in \mathcal{L}(\hilb_\tsc{a}\otimes\hilb_\tsc{b})$} can simultaneously satisfy Eqs.~\eqref{Eq: marginal expectation values} and~\eqref{Eq: correlations}. Notably, this issue would also arise with a past-light-cone update. More generally, even if the update is performed non-covariantly with respect to a specific foliation, there will always exist a spacelike hypersurface where Eq.~\eqref{Eq: correlations} is not satisfied.

In fact, the inability to satisfy Eq.~\eqref{Eq: correlations} across all possible spacelike hypersurfaces lies at the heart of the criticisms of state updates in~\cite{AharonovAlbert1980}, one of which was that they ultimately lead to violations of the conservation of global charges: For instance, if in the previous setup A and B are holding boxes, and $\ket{1_j}$ and $\ket{0_j}$ represent the number of electrons in each box, the local charge is $\hat q_j = (\hat\sigma_{z,j} - \mathds{1}_j)/2$, and the total charge is $\hat Q = \hat q_\tsc{a} + \hat q_\tsc{b}$. Then, if the update is performed along the future light cone, Eq.~\eqref{Eq: marginal expectation values} implies that across any hypersurface containing $\msf x_\tsc{a}(\tau_\tsc{a}^{**})$ and $\msf x_\tsc{b}(\tau_\tsc{b}^*)$ we will have $\langle \hat Q \rangle = \langle \hat q_\tsc{a} \rangle + \langle \hat q_\tsc{b} \rangle = 0 -\tfrac{1}{2} \neq -1$. Moreover, this is not only a problem of the future lightcone update: since Eq.~\eqref{Eq: correlations} is not satisfied in general for any prescription of the update rule, there will always be hypersurfaces across which $\langle \hat Q \rangle \neq -1$, i.e., the total charge is not conserved.

The discussion above shows that a representation of multipartite systems in terms of tensor products does not leave room for a covariant update rule that 
\begin{itemize}
\item[$(i)$] is \textit{fully predictive}, i.e., incorporates knowledge of the measurement outcome into the description of the multipartite system, accounts for correlations, and ensures compatibility between sequential measurements applied to both the joint system and its subsystems, 
\item[$(ii)$] \textit{respects ignorance}\footnote{This wording is inspired by C. J. Fewster’s ``protection of ignorance'', also known as the \textit{principle of blissful ignorance}~\cite{Fewster2025lectures}, which, albeit different in meaning, is similar in spirit to the notion presented here.}, i.e., incorporates that the propagation of information gathered from measurements is constrained by the causal structure of spacetime, reflecting how an observer's local state of knowledge depends on their location.
\end{itemize}
        
Our goal is to build a formalism that accommodates (\textit{i}) and (\textit{ii}) to allow for a covariant, fully predictive update rule that respects ignorance and is free from the problem of conservation of global charges identified by Aharonov and Albert.

\noindent\textit{\textbf{Polyperspective formalism.---}} Prescribing an update rule that is both \textit{fully predictive} and \textit{respects ignorance} requires addressing the apparent contradiction posed by Eqs.~\eqref{Eq: marginal expectation values} and~\eqref{Eq: correlations}. To resolve this we  distinguish between the prediction of outcomes of measurements performed independently on $\text{A \textit{or} B}$, and the prediction of outcomes of experiments performed jointly on $\text{A \textit{and} B}$. Namely, we call \textit{individual observables} those that can be measured independently on A or B, and their expectations can be computed using \textit{local} states $\rhoh_\tsc{a}$ and $\rhoh_\tsc{b}$.  Meanwhile, \textit{joint observables} are those for which predicting their outcomes requires knowledge of the composite system $\tsc{A}\tsc{B}$, and computing their expectations requires a joint state $\rhoh_\tsc{ab}$.

Implementing this distinction requires considering that $\hat A \in \mathcal{L}(\hilb_\tsc{a})$ (as an \textit{individual} observable) and \mbox{$\hat A \otimes \mathds{1}_\tsc{b} \in \mathcal{L}(\hilb_\tsc{a} \otimes \hilb_\tsc{b})$} (as a \textit{joint} observable) are different operators, even though $\hat A \otimes \mathds{1}_\tsc{b}$ acts trivially on the B sector of the joint system. That is, measuring an observable of $\tsc{A}$ while having access to information from $\tsc{B}$ is different from measuring the same observable without knowledge about $\tsc{B}$. A convenient way to make this difference explicit is to consider an extended joint Hilbert space
\begin{equation}\label{Eq: extended Hilbert}
    \tilde{\hilb}_\tsc{ab} \coloneqq \hilb_{\tsc{a}} \oplus \hilb_{\tsc{b}} \oplus (\hilb_{\tsc{a}}\otimes\hilb_{\tsc{b}}),
\end{equation}
with the space of physical operators is given by\footnote{Each summand in Eq.~\eqref{Eq: extended Hilbert} formally acts as a superselection sector.}
\begin{equation}
    \mathcal{L}(\tilde\hilb_\tsc{ab})_\text{phys} \coloneqq \mathcal{L}(\hilb_\tsc{a}) \oplus \mathcal{L}(\hilb_\tsc{b}) \oplus \mathcal{L}(\hilb_\tsc{a} \otimes \hilb_\tsc{b}).
\end{equation}
This structure admits a straightforward intuitive interpretation: $\mathcal{L}(\hilb_j)$ is the space of \textit{individual observables} of \mbox{$j\in\{A,B\}$}, while $\mathcal{L}(\hilb_\tsc{a} \otimes \hilb_\tsc{b})$ is the space of joint observables. The full state of the system is described by an operator of the form
\begin{equation}
    \tilde{\rho}_\tsc{ab} \coloneqq \rhoh_\tsc{a} \oplus \rhoh_\tsc{b} \oplus \rhoh_\tsc{ab} \in \mathcal{L}(\tilde\hilb_\tsc{ab})_\text{phys},
\end{equation}
which we call \textit{polyperspective state} (\textit{polystate} for short). In general, $\text{the density operators } \rhoh_\tsc{a} \text{ and } \rhoh_\tsc{b}$ will \textit{not} coincide with the corresponding partial traces\footnote{E.g., given individual observables $\hat A \in \mathcal{L}(\hilb_\tsc{a})$ and $\hat B \in \mathcal{L}(\hilb_\tsc{b})$, and a joint observable $\hat C \in \mathcal{L}(\hilb_\tsc{a}\otimes\hilb_\tsc{b})$, their expectation values can be computed from $\tilde\rho$ as \mbox{$\langle \hat A \rangle_{\tilde\rho} =\Tr\normalsize(\tilde{\rho}_\tsc{ab} \hat A\normalsize) =\Tr_\tsc{a}(\rhoh_\tsc{a}\hat A)$}, \mbox{$\langle \hat B \rangle_{\tilde\rho} = \Tr_\tsc{b}(\rhoh_\tsc{b}\hat B)$}, \mbox{$\langle \hat C \rangle_{\tilde\rho} = \Tr_\tsc{ab}(\rhoh_\tsc{ab}\hat C)$}.} of $\rhoh_\tsc{ab}$. For a generalization to $n$ subsystems see App.~\ref{Appendix: generalization to n subsystems}.

The formalism leaves room to incorporate the dependence of $\tilde\rho$ on two time parameters, with the proper times $\tau_\tsc{a}$ and $\tau_\tsc{b}$ being natural choices. This removes the need for a preferred foliation to parametrize the system's time evolution, making it particularly convenient for measurement updates along lightcones, which is at odds with a unique description of the quantum state across all hypersurfaces of a spacelike foliation. In general, the state $\tilde\rho(\tau_\tsc{a},\tau_\tsc{b})$ can be written as
\begin{equation}\label{Eq: rho tilde depends on two time parameters}
    \tilde\rho(\tau_\tsc{a},\tau_\tsc{b}) = \rhoh_\tsc{a}(\tau_\tsc{a}) \oplus \rhoh_\tsc{b}(\tau_\tsc{b}) \oplus \rhoh_\tsc{ab}(\tau_\tsc{a},\tau_\tsc{b}).
\end{equation}
Notice that the local states $\rhoh_\tsc{a}$ and $\rhoh_\tsc{b}$ only depend on $\tau_\tsc{a}$ and $\tau_\tsc{b}$, respectively. This is because they yield expectation values of individual observables, which depend solely on local knowledge of $\text{A or B}$ and should therefore rely only on the respective proper time at which the measurement is performed.

In general, to find $\tilde\rho(\tau_\tsc{a},\tau_\tsc{b})$, we define the completely positive map
\begin{equation}\label{Eq: Psi maps operations in spacetime}
    \Psi_{\mathcal S}:\mathcal{L}(\hilb_\tsc{a}\otimes\hilb_\tsc{b}) \to \mathcal{L}(\hilb_\tsc{a}\otimes\hilb_\tsc{b}) 
\end{equation}
that encodes all the transformations undergone by $\text{A and B}$ in $\mathcal S \subset \mathcal{M}$. This includes both time evolution and measurements. Since we prescribed that updates should be implemented along the future lightcone of the measurement event, the only transformations that are relevant to determine the local states $\rhoh_\tsc{a}(\tau_\tsc{a})$ and $\rhoh_\tsc{b}(\tau_\tsc{b})$ lie in the causal past of $\msf x_\tsc{a}(\tau_\tsc{a})$ and $\msf x_\tsc{b}(\tau_\tsc{b})$, respectively. Specifically, given an initial state of the system $\rhoh_\tsc{ab}$, and its associated polystate
\begin{equation}
\tilde\rho = \rhoh_\tsc{a} \oplus \rhoh_\tsc{b} \oplus \rhoh_\tsc{ab},
\end{equation}
where $\hat{\rho}_\tc{a} = \Tr_\tc{b}(\hat{\rho}_\tc{ab})$ and $\hat{\rho}_\tc{b} = \Tr_\tc{a}(\hat{\rho}_\tc{ab})$, we define 
\begin{align}
    \rhoh_\tsc{a}(\tau_\tsc{a}) & \propto \Tr_\tsc{b}\Psi_{\mathcal{J}^-(\msf x_\tsc{a}(\tau_\tsc{a}))}(\rhoh_\tsc{ab}), \label{Eq: rhoh_A tau_a}\\
    \rhoh_\tsc{b}(\tau_\tsc{b}) & \propto \Tr_\tsc{a}\Psi_{\mathcal{J}^-(\msf x_\tsc{b}(\tau_\tsc{b}))}(\rhoh_\tsc{ab}), \label{Eq: rhoh_B tau_b}
\end{align}
where $\mathcal{J}^-(\msf x)$ denotes the causal past of $\msf x \in \mathcal M$, and the proportionality constants are determined by the normalization of $\rhoh_\tsc{a}(\tau_\tsc{a})\text{ and }\rhoh_\tsc{b}(\tau_\tsc{b})$.

To complete the description of $\tilde\rho(\tau_\tsc{a},\tau_\tsc{b})$ for arbitrary $(\tau_\tsc{a},\tau_\tsc{b})$, we need to obtain $\rhoh_\tsc{ab}(\tau_\tsc{a},\tau_\tsc{b})$, i.e., the density operator used to compute expectations of joint observables. The joint state assigned to A and B should account only for transformations within \textit{both} of their causal pasts, therefore 
\begin{equation}\label{Eq: rhoh_AB tau_a,tau_b}
    \rhoh_\tsc{ab} (\tau_\tsc{a},\tau_\tsc{b}) \propto \Psi_{\mathcal{J}^-(\msf x_\tsc{a}(\tau_\tsc{a})) \,\cup\, \mathcal{J}^-(\msf x_\tsc{b}(\tau_\tsc{b}))}(\rhoh_\tsc{ab}).
\end{equation}
Measuring joint observables requires performing measurements on $\text{A and B}$ separately and aggregating the results in a spacetime region that must lie in the future of both operations---which we call the \textit{processing region}~\cite{MeasurementTheory,Ruep2021}. Given a processing region $\mathcal{P}$, the associated joint state should account only for the transformations within $\mathcal J^-(\mathcal P)$. With the prescription~\eqref{Eq: rhoh_AB tau_a,tau_b}, $\rhoh_\tsc{ab}(\tau_\tsc{a},\tau_\tsc{b})$ encodes information about the transformations that \textit{any} processing region will have access to\footnote{In fact, for certain spacetimes, it encodes the \textit{maximum} information that \textit{all} processing regions will have access to. The `maximum information that all processing regions will have access to' corresponds to the information contained in their \textit{common past}, which in this case is given by $\bigcap \big\{\mathcal J^-(\msf y) \, : \, \msf y \in \mathcal{J}^+(\msf x_\tsc{a}(\tau_\tsc{a})) \cap \mathcal{J}^+(\msf x_\tsc{b}(\tau_\tsc{b}))\big\}$,
i.e., the common past of all points in the intersection of the causal futures of $\msf x_\tsc{a}(\tau_\tsc{a})$ and $\msf x_\tsc{b}(\tau_\tsc{b})$. While this set will generally be strictly larger than $\mathcal{J}^-(\msf x_\tsc{a}(\tau_\tsc{a})) \cup \mathcal{J}^-(\msf x_\tsc{b}(\tau_\tsc{b}))$, the two sets will coincide in some special spacetimes, such as the Lorentzian cylinder $\R{} \times \mathbb{S}^1$.}.

In the Bell pair example, this formalism naturally accommodates the predictions of an update rule that applies only in the future of selective measurements. The scenario where Alice measures $+1 \text{ at } \tau_\tc{a} = \tau_\tc{a}^*$ is encoded in
\begin{equation}\label{Eq: transformations map case by case}
    \Psi_\mathcal{S}(\rhoh) = \left\{ \begin{array}{lr}
    \rhoh & \text{if   } \msf x_\tsc{a}(\tau_\tsc{a}^*) \notin \mathcal S, \\[2mm]
    \proj{0_\tsc{a}}{0_\tsc{a}} \,\rhoh\, \proj{0_\tsc{a}}{0_\tsc{a}} \; & \text{if   } \msf x_\tsc{a}(\tau_\tsc{a}^*) \in \mathcal S.
    \end{array} \right.
\end{equation}\\
Hence, if the system is initially in the Bell state $\ket{\Psi^+}$, Eqs.~\eqref{Eq: rhoh_A tau_a}--\eqref{Eq: rhoh_AB tau_a,tau_b} imply that the polystate at the time of the measurement $\tau_\tsc{a}^*$ is
\begin{equation}\label{Eq: rho tilde Bell pair case}
    \tilde\rho (\tau_\tsc{a}^*,\tau_\tsc{b}) = \proj{0_\tsc{a}}{0_\tsc{a}}\oplus \tfrac{1}{2}\mathds{1}_\tsc{b}\oplus \proj{0_\tsc{a}1_\tsc{b}}{0_\tsc{a}1_\tsc{b}},
\end{equation}
for any $\tau_\tsc{b} \in (\tau_\tsc{b}^-,\tau_\tsc{b}^+)$. This state correctly reproduces the expectation values in Eqs.~\eqref{Eq: marginal expectation values}--\eqref{Eq: correlations} and, in particular, encodes the prescription for $\rhoh_\tsc{a}$, $\rhoh_\tsc{b}$ and $\rhoh_\tsc{ab}$ given in~\cite{MeasurementTheory}. 

The recipe given by Eqs.~\eqref{Eq: rhoh_A tau_a}--\eqref{Eq: rhoh_AB tau_a,tau_b} is general: it is independent of the specific transformations undergone by the system. In particular, it can accommodate freedom of choice of measurement in an EPR test (see App.~\ref{Appendix: Update Bell Pair}).


\noindent\textit{\textbf{Statistics of qubits in spacetime.---}} The use of polystates to describe the full joint state of quantum systems in spacetime does not change the interpretation of single-shot experiments. However, it admits a particularly clean statistical interpretation.

The  statistical interpretation of quantum mechanics asserts that the state describes the statistical properties of an ensemble of identically prepared systems~\cite{Ballentine1970}. While it relinquishes the ambition of an ontological description of states, it has the advantage of offering an operational interpretation for selective updates as post-selection: realizations that do not yield the prescribed outcome are discarded.

To illustrate this, we revisit the Bell pair scenario: let Alice and Bob follow  timelike trajectories $\msf x_\tsc{a}(\tau_\tsc{a})$ and $\msf x_\tsc{b}(\tau_\tsc{b})$ while sharing $N$ pairs of qubits initially prepared in the Bell state $\ket{\Psi^+}$. At $\tau_\tsc{a}=\tau_\tsc{a}^*$, Alice measures $\hat\sigma_z$ on the qubits she holds, and keeps only the subset of them for which the outcome was $+1$. The ensemble of qubits she retains is described by the density operator $\proj{0_\tsc{a}}{0_\tsc{a}}$, precisely matching $\rhoh_\tsc{a}(\tau_\tsc{a}^*)$ in Eq.~\eqref{Eq: rho tilde Bell pair case}. Meanwhile, at any time before $\tau_\tsc{b}^+$, and without performing any measurements on his qubits, Bob cannot know which pairs were selected or discarded by Alice. Until Bob receives information from Alice at $\tau_\tsc{b} \geq \tau_\tsc{b}^+$ regarding which qubits to discard, he must hold on to all $N$ of them. Consequently, his ensemble is described by $\mathds{1}_\tsc{b}/2$. This is precisely $\rhoh_\tsc{b}(\tau_\tsc{b})$ in Eq.~\eqref{Eq: rho tilde Bell pair case}, for any $\tau_\tsc{b} \in (\tau_\tsc{b}^-,\tau_\tsc{b}^+)$. Finally, to measure correlations between their subsystems, Alice and Bob must consider only the pairs that remain complete (since Alice discarded the qubits that returned $-1$, even though Bob retained their counterparts). At any point where their results can be combined (i.e., in any processing region), the non-discarded pairs will display the appropriate correlations, described by $\hat\rho_\tsc{ab}=\proj{0_\tsc{a}1_\tsc{b}}{0_\tsc{a}1_\tsc{b}}$, which coincides with $\rhoh_\tsc{ab}(\tau_\tsc{a}^*,\tau_\tsc{b})$ in Eq.~\eqref{Eq: rho tilde Bell pair case}.

\noindent\textit{\textbf{Observer-dependent frameworks.---}} This formalism can be related with an observer-dependent framework, which naturally applies to spatially extended systems and quantum fields.

To  see this,  notice that from A's perspective, the joint state of the system at time $\tau_\tsc{a}$ is proportional to $\Psi_{\mathcal J^-(\msf x_\tsc{a}(\tau_\tsc{a}))}(\rhoh_\tsc{ab})$, and the partial trace of \textit{this} joint state yields $\rhoh_\tsc{a}(\tau_\tsc{a})$ (which, in general, will not coincide with the partial trace of $\rhoh_\tsc{ab}(\tau_\tsc{a},\tau_\tsc{b})$, as Eq.~\eqref{Eq: rho tilde Bell pair case} exemplifies). The same applies to B.

More generally, given an initial state $\rhoh_0$, we can define
\begin{equation}\label{Eq: spacetime-dependent state}
    \rhoh(\msf x) \coloneqq \frac{\Psi_{\mathcal J^-(\msf x)}(\rhoh_0)}{\Tr\big[ \Psi_{\mathcal J^-(\msf x)}(\rhoh_0) \big]},
\end{equation}
which is the state of the system as described by a maximally-informed observer~\cite{Pranzini2023} at $\msf x \in \spt$. This generalization of the formalism is particularly convenient for spatially extended non-relativistic systems and quantum field theories~\cite{MeasurementTheory, MeasurementTheoryUpdated}. 

This observer-dependent approach allows us to recover a description of the system's evolution in terms of \textit{one} time parameter, as in non-relativistic quantum mechanics. The fundamental difference lies in that this account of the evolution depends entirely on the observer that recalls it, or, more concretely, on their worldline: given a timelike trajectory $\msf z(\tau)$ parametrized by its proper time $\tau$, from Eq.~\eqref{Eq: spacetime-dependent state} we can define its associated \textit{recollection} as
\begin{equation}\label{Eq: trajectory-dependent evolution}
    \rhoh_\msf{z}(\tau) \coloneqq \rhoh(\msf z(\tau)) \propto \Psi_{\mathcal J^-(\msf z(\tau))}(\rhoh_0),
\end{equation}
where $\rhoh_0$ is the initial state of the system\footnote{For simplicity, we assume that the system has always existed, and that $\rhoh_0 = \lim_{\tau \to -\infty} \rhoh(\tau)$. This assumption can be easily lifted by considering that the system was (jointly) prepared at some point in spacetime, $\msf x_0$, and therefore its history as recalled from the trajectory $\msf z(\tau)$ would start at some $\tau_0$, which corresponds with the intersection of the future light cone of $\msf x_0$ with the worldline $\msf z(\tau)$.}.

\noindent\textit{\textbf{Covariance and conservation of charges.---}} Finally, we examine whether the obstructions identified in~\cite{AharonovAlbert1980} to defining a well-behaved state in relativistic quantum mechanics are mitigated within the polyperspective formalism. 

First, note that the polystate is covariant by construction: it is defined solely from the causal structure of spacetime and is frame-independent. The same applies to $\rhoh(\msf x)$ (Eq.~\eqref{Eq: spacetime-dependent state}). 

Second, let $\{\Sigma_t\}_{t\in\R{}}$ be a foliation of $\mathcal M$, associated with a global time function $t$, and consider a bipartite system. Denoting with $\tau_\tsc{a}(t)$ and $\tau_\tsc{b}(t)$ the proper times at which A’s and B’s worldlines intersect the leaf $\Sigma_t$, we identify the state of the system on $\Sigma_t$ as $\tilde\rho(\tau_\tsc{a}(t),\tau_\tsc{b}(t))$. To make predictions for observables that are non-local, like total charge, involves computing expectations using all the information available on each leaf, i.e., using the joint state, $\rhoh_\tsc{ab}(\tau_\tsc{a}(t),\tau_\tsc{b}(t))$. In the observer-dependent framework, the state within the foliation is
\begin{equation}\label{Eq: foliation-dependent state}
    \rhoh_\Sigma(t) \coloneqq \frac{\Psi_{\mathcal J^-(\Sigma_t)}(\rhoh_0)}{\Tr\big[ \Psi_{\mathcal J^-(\Sigma_t)}(\rhoh_0) \big]}. 
\end{equation}
Notice that while the polystate $\tilde\rho$ is covariant, the foliation-dependent state $\rhoh_\Sigma$, as given by Eq.~\eqref{Eq: foliation-dependent state}, depends explicitly on the choice of foliation $\{\Sigma_t\}_{t\in\R{}}$ (see App.~\ref{Appendix: Cauchy}). This distinction between $\tilde\rho$ and $\rhoh_\Sigma$ highlights the key difference between the polyperspective framework and frame-dependent formulations~\cite{Malin1982, Glowacki2024}, which in general rely on non-local information even when computing expectation values of local observables.

The violation of the conservation of electric charge identified in~\cite{AharonovAlbert1980} as an unavoidable byproduct of collapse-like state updates is resolved in the polyperspective formalism by realizing that the charge density $\hat q(\msf x)$ is a \textit{local} operator, while the total charge of the system in a particular leaf, \mbox{$\hat Q(t) = \int_{\Sigma_t} \!\dd\Sigma \; \hat q(\msf x)$}, is a \textit{non-local} operator\footnote{This was explicitly acknowledged by Aharonov and Albert, who clarified in a footnote: ``The charge, that is, measured nonlocally''~\cite{AharonovAlbert1980}.}. As a consequence, in this formalism the density operators used to compute the expectation values $\langle \hat q(\msf x)\rangle$, and the one used to compute $\langle \hat Q(t) \rangle$, are generically different, i.e., in general,
\begin{equation}
\bigg\langle \int_{\Sigma_t} \dd \Sigma\, \hat q ( \msf x) \bigg\rangle_{\!\!\tilde\rho} \neq \int_{\Sigma_t} \dd \Sigma \, \langle \hat q (\msf x) \rangle_{\tilde\rho}\,\tgg{,} 
\end{equation}
where one should use $\hat{\rho}_\Sigma(t)$ (the joint sector of the polystate) on the left-hand side, and $\hat{\rho}(\msf x)$ (the individual sectors) on the right-hand side.

Furthermore, the set of all recollections \{$\rhoh_\msf{z}(\tau)\}$ encompasses all the possible accounts of the evolution of the system. Regardless of what we consider to be the intrinsic description of the state, everything that any observer can access is encoded in one of these recollections. In particular, the expectation value of the total charge, $\langle \hat Q (t) \rangle_{\mathsf z} \equiv \Tr\normalsize[\rhoh_\msf{z}(\tau(t)) \hat Q(t)\normalsize]$ is conserved for all $\msf z (\tau)$. 

\noindent\textit{\textbf{Conclusions.---}} We have addressed the longstanding problem of how to consistently update the state of multipartite quantum systems after measurements have been performed in relativistic settings. This involved balancing two central demands: (i) compatibility with sequential local and joint measurements, and (ii) the constraints imposed by causality on the propagation of information gathered from measurements.

We demonstrated that state updates constrained to the future lightcones of measurements encode the causal propagation of information but fail to preserve multipartite correlations. In contrast, updates along the past lightcone---like those proposed in earlier literature---implicitly entail retroactive updates that either compromise the interpretation of the quantum state as evolving under physical processes in spacetime, or fail to acknowledge where in spacetime information is available. Importantly, both prescriptions lead to violations of charge conservation.

To resolve this tension we proposed a framework where the quantum state is described not by a single, but by a multiplet of density operators, each of which encodes the information available to different subsystems. This construction allows individual and joint observables to be treated distinctly, enabling causal updates that preserve both local and global statistical predictions.

This formalism admits a natural reinterpretation in terms of observer-dependent states, yielding a covariant description of quantum information flow. It ensures compatibility with relativistic causality and---unlike previous proposals---respects conservation laws across arbitrary spacetime foliations.

Altogether, this work provides a fully predictive and relativistically self-consistent framework for describing selective measurements in relativistic quantum information, with direct implications for foundational questions and practical quantum information protocols---such as those involving entanglement in quantum field theories, and distributed quantum systems in curved spacetimes.

\acknowledgements   

This work was partially inspired by Fay Dowker's \textit{Useless Qubits in ``Relativistic Quantum Information''}~\cite{Dowker2011Useless}, where the author pointed out the salient issues of the measurement problem in relativistic setups to the RQI community. We are also grateful for the invitation to the GRASP-Q Workshop organized by the Instituto de Ciencias Nucleares at UNAM, where we had the chance to have enlightening discussions about these topics with Benito A. Ju\'arez-Aubry, Daniel Sudarsky, and El\'ias Okon. We would like to thank the Quantum Fields and Gravity group for their comments and insights, as well as Christopher J. Fewster, Ana Alonso-Serrano, Nicola Pranzini, and Esko Keski-Vakkuri for helpful discussions, and Luis J. Garay and Roope Uola for comments on an earlier version of this draft. JPG acknowledges the support of a Mike and Ophelia Lazaridis Fellowship, as well as the support of a fellowship from ``La Caixa'' Foundation (ID 100010434, code LCF/BQ/AA20/11820043). TRP is thankful for financial support from the Olle Engkvist Foundation (no.225-0062). This work was partially conducted while TRP was still a PhD student at the Department of Applied Mathematics at the University of Waterloo, the Institute for Quantum Computing, and the Perimeter Institute for Theoretical Physics. TRP acknowledges partial support from the Natural Sciences and Engineering Research Council of Canada (NSERC) through the Vanier Canada Graduate Scholarship. EMM acknowledges support through the Discovery Grant Program of the Natural Sciences and Engineering Research Council of Canada (NSERC).  EMM also acknowledges support of his Ontario Early Researcher award. Research at Perimeter Institute is supported in part by the Government of Canada through the Department of Innovation, Science and Industry Canada and by the Province of Ontario through the Ministry of Colleges and Universities. Nordita is partially supported by Nordforsk.

	
\onecolumngrid

\appendix

\section{The polyperspective formalism for arbitrary composite systems}\label{Appendix: generalization to n subsystems}

In the main text, we introduced the polyperspective formalism for a composite system with two subsystems A and B (see Eqs.~\eqref{Eq: extended Hilbert}--\eqref{Eq: rho tilde depends on two time parameters}). It is straightforward to generalize this construction to an arbitrary composite system with $n$ subsystems. Namely, consider $n$ non-relativistic systems following timelike trajectories $\msf x_i(\tau_i)$, parametrized by their proper times $\tau_i$, for \mbox{$i\in\{1,\hdots,n\}$}. The associated extended joint Hilbert space is 
\begin{equation}
\tilde H \coloneqq \bigoplus_{i} \hilb_i \oplus \bigoplus_{i<j} (\hilb_i \otimes \hilb_j) \oplus \bigoplus_{i<j<k} (\hilb_i \otimes \hilb_j \otimes \hilb_k) \oplus \hdots \oplus (\hilb_1 \otimes \hdots \otimes \hilb_n),
\end{equation}
which is a direct sum of all the possible tensor products of $m$ local Hilbert spaces, for every $m \in \{1,\hdots,n\}$. The space of physical operators is then given by
\begin{align}
    \mathcal L(\tilde H)_\text{phys} \coloneqq \bigoplus_{i}\mathcal{L}(\hilb_i) \oplus \bigoplus_{i<j} \mathcal{L} (\hilb_i \otimes \hilb_j) \oplus \bigoplus_{i<j<k} \mathcal{L} (\hilb_i \otimes \hilb_j \otimes \hilb_k) \oplus \hdots \oplus \mathcal{L}(\hilb_1 \otimes \hdots \otimes \hilb_n).
\end{align}
The joint system is described by the polyperspective state 
\begin{align}\label{Eq: rho tilde general}
    \tilde\rho(\tau_1,\hdots,\tau_n) = \bigoplus_i \rhoh_i(\tau_i) \oplus \bigoplus_{i<j} \rhoh_{ij}(\tau_i,\tau_j) \oplus \bigoplus_{i<j<k} \rhoh_{ijk}(\tau_i,\tau_j,\tau_k) \oplus \hdots \oplus \rhoh_{1\hdots n}(\tau_1,\hdots,\tau_n),
\end{align}
where, for any ordered subset of indices $I \subset \{1,\hdots,n\}$, 
\begin{equation}\label{Eq: parts of rho tilde general}
\rhoh_I (\{\tau_i\,:\, i\in I\}) \propto \Tr_{I^c} \Psi_{\bigcup_{i\in I} \mathcal J^-(\msf x_i (\tau_i))}(\rhoh_{1\hdots n}), 
\end{equation}
which generalizes Eqs.~\eqref{Eq: rhoh_A tau_a}--\eqref{Eq: rhoh_AB tau_a,tau_b}. Here, $\rhoh_{1\hdots n}$ is the initial joint state, and $I^c$ is the complementary set of $I$ in $\{1,\hdots,n\}$. As before (see Eq.~\eqref{Eq: Psi maps operations in spacetime}), $\Psi_{\mathcal S}: \mathcal L(\hilb_1 \otimes \hdots \otimes \hilb_n) \to \mathcal L(\hilb_1 \otimes \hdots \otimes \hilb_n)$ encodes all the transformations undergone by the subsystems of the composite system in the causally convex set $\mathcal S \subset \mathcal M$.

\section{Further updating a Bell pair in the polyperspective formalism}\label{Appendix: Update Bell Pair}

In the main text, we considered a setup where two qubits, A and B, following timelike trajectories $\msf{x}_\tsc{a}(\tau_\tsc{a})$ and $\msf{x}_\tsc{b}(\tau_\tsc{b})$, are initially in the Bell state $\ket{\Psi^+} = (\ket{0_\tsc{a}}\ket{1_\tsc{b}} + \ket{1_\tsc{a}}\ket{0_\tsc{b}})/\sqrt{2}$. When Alice measures $\hat\sigma_{z}$ at $\tau_\tsc{a} = \tau_\tsc{a}^*$ and obtains a $+1$ outcome, the spacetime-dependent maps encoding this transformation are given by Eq.~\eqref{Eq: transformations map case by case},
\begin{equation}\label{Eq: transformations map case by case bis}
    \Psi_\mathcal{S}(\rhoh) = \left\{ \begin{array}{lr}
    \rhoh & \text{if   } \msf x_\tsc{a}(\tau_\tsc{a}^*) \notin \mathcal S, \\[2mm]
    \proj{0_\tsc{a}}{0_\tsc{a}} \,\rhoh\, \proj{0_\tsc{a}}{0_\tsc{a}} \; & \text{if   } \msf x_\tsc{a}(\tau_\tsc{a}^*) \in \mathcal S.
    \end{array} \right.
\end{equation}\\
From Eqs.~\eqref{Eq: rhoh_A tau_a}--\eqref{Eq: rhoh_AB tau_a,tau_b}, the complete description of the (time-dependent) polystate is given by
\begin{equation}\label{Eq: time-dependent polystate sigma_z}
    \tilde\rho(\tau_\tsc{a},\tau_\tsc{b}) = \left\{ \begin{array}{lr}
    \tfrac{1}{2}\mathds{1}_\tsc{a}\oplus \tfrac{1}{2}\mathds{1}_\tsc{b}\oplus \proj{\Psi^+}{\Psi^+} & \text{if   } \tau_\tsc{a} < \tau_\tsc{a}^*, \tau_\tsc{b} < \tau_\tsc{b}^+, \\[2mm]
    \proj{0_\tsc{a}}{0_\tsc{a}}\oplus \tfrac{1}{2}\mathds{1}_\tsc{b}\oplus \proj{0_\tsc{a}1_\tsc{b}}{0_\tsc{a}1_\tsc{b}} & \text{if   } \tau_\tsc{a} \geq \tau_\tsc{a}^*, \tau_\tsc{b} < \tau_\tsc{b}^+, \\[2mm]
    \tfrac{1}{2}\mathds{1}_\tsc{a} \oplus \proj{1_\tsc{b}}{1_\tsc{b}}\oplus \proj{0_\tsc{a}1_\tsc{b}}{0_\tsc{a}1_\tsc{b}} & \text{if   } \tau_\tsc{a} < \tau_\tsc{a}^*, \tau_\tsc{b} \geq \tau_\tsc{b}^+, \\[2mm]
    \proj{0_\tsc{a}}{0_\tsc{a}}\oplus \proj{1_\tsc{b}}{1_\tsc{b}}\oplus \proj{0_\tsc{a}1_\tsc{b}}{0_\tsc{a}1_\tsc{b}} & \text{if   } \tau_\tsc{a} \geq \tau_\tsc{a}^*, \tau_\tsc{b} \geq \tau_\tsc{b}^+.
    \end{array} \right.
\end{equation}\\
Of course, the applicability of the polyperspective formalism is not contingent on Alice's outcome (we could proceed similarly if the outcome had been $-1$, by replacing every $0$ with a $1$ and vice versa), nor on the observable that she measures. For instance, if Alice measures $\hat\sigma_x$ instead of $\hat\sigma_z$, obtaining a $\pm 1$ outcome, the resulting polystate as a function of the proper times $\tau_\tsc{a}$ and $\tau_\tsc{b}$ is again given by Eq.~\eqref{Eq: time-dependent polystate sigma_z}, but replacing every $0$ and $1$ with $\pm$, where \mbox{$\ket{\pm_j} = (\ket{0_j} \pm \ket{1_j})/\sqrt{2}$} are the eigenstates of $\hat\sigma_{x,j}$, for $j \in \{\tsc{A},\tsc{B}\}$. 

Indeed, as pinpointed in the main text, the recipe given by Eqs.~\eqref{Eq: rhoh_A tau_a}--\eqref{Eq: rhoh_AB tau_a,tau_b} (extended to arbitrary composite systems in Eq.~\eqref{Eq: parts of rho tilde general}), is general, i.e., it does not depend on the initial state of the system nor on the specific transformations and measurements performed on it. Applying it to each concrete scenario amounts to identifying the appropriate family of maps $\Psi_\mathcal{S}$ for that scenario, and using Eq.~\eqref{Eq: parts of rho tilde general} to obtain the polystate $\tilde\rho$ as a function of the proper times parametrizing the trajectories of the subsystems. 

One particularly relevant application of the polyperspective formalism is the analysis of an EPR test. To make things concrete, let us consider the initial joint state of the qubits to be the singlet state
\begin{equation}
    \ket{\Psi^-} = \frac{1}{\sqrt{2}}(\ket{0_\tsc{a}}\ket{1_\tsc{b}} - \ket{1_\tsc{a}}\ket{0_\tsc{b}}),
\end{equation}
which it is worth recalling is rotationally invariant. We consider that Alice still measures $\hat\sigma_z$ at $\tau_\tsc{a} = \tau_\tsc{a}^*$, but now Bob also measures $\hat\sigma_{\bm n}$ at $\tau_\tsc{b} = \tau_\tsc{b}^*$, where we denote with $\theta_\tsc{ab}$ the angle between $\bm n$ and the $z$-axis, and $\tau_\tsc{b}^*$ is such that $\msf x_\tsc{a}(\tau_\tsc{a}^*)$ and $\msf x_\tsc{b}(\tau_\tsc{b}^*)$ are spacelike separated. Let $a$ and $b$ be the outcomes that Alice and Bob obtain in their respective measurements, and let us denote with $\ket{a}$ and $\ket{b}$ (and $\ket{-a}$ and $\ket{-b}$) the corresponding eigenstates (and their orthonormal counterparts) of $\hat\sigma_{z}$ and $\hat\sigma_{\bm n}$, respectively. The spacetime-dependent maps encoding this scenario are given by
\begin{equation}\label{Eq: transformations map case by case EPR}
    \Psi_\mathcal{S}(\rhoh) = \left\{ \begin{array}{ll}
    \rhoh & \text{if   } \msf x_\tsc{a}(\tau_\tsc{a}^*), \msf x_\tsc{b}(\tau_\tsc{b}^*) \notin \mathcal S, \\[2mm]
    \proj{a_\tsc{a}}{a_\tsc{a}}\,\rhoh\,\proj{a_\tsc{a}}{a_\tsc{a}} & \text{if   } \msf x_\tsc{a}(\tau_\tsc{a}^*) \in \mathcal S, \msf x_\tsc{b}(\tau_\tsc{b}^*) \notin \mathcal S, \\[2mm]
    \proj{b_\tsc{b}}{b_\tsc{b}}\,\rhoh\,\proj{b_\tsc{b}}{b_\tsc{b}} & \text{if   } \msf x_\tsc{a}(\tau_\tsc{a}^*) \notin \mathcal S, \msf x_\tsc{b}(\tau_\tsc{b}^*) \in \mathcal S, \\[2mm]
    \ket{a_\tsc{a}}\!\ket{b_\tsc{b}}\!\! \bra{a_\tsc{a}}\!\bra{b_\tsc{b}} \rhoh \ket{a_\tsc{a}}\!\ket{b_\tsc{b}}\!\! \bra{a_\tsc{a}}\!\bra{b_\tsc{b}} \; & \text{if   } \msf x_\tsc{a}(\tau_\tsc{a}^*), \msf x_\tsc{b}(\tau_\tsc{b}^*) \in \mathcal S.
    \end{array} \right.
\end{equation}\\
As before, from Eqs.~\eqref{Eq: rhoh_A tau_a}--\eqref{Eq: rhoh_AB tau_a,tau_b}, the complete description of the (time-dependent) polystate is given by
\begin{equation}\label{Eq: time-dependent polystate EPR}
    \tilde\rho_{ab}(\tau_\tsc{a},\tau_\tsc{b}) = \left\{ \begin{array}{lr}
    \tfrac{1}{2}\mathds{1}_\tsc{a}\oplus \tfrac{1}{2}\mathds{1}_\tsc{b}\oplus \proj{\Psi^-}{\Psi^-} & \text{if   } \tau_\tsc{a} < \tau_\tsc{a}^*, \tau_\tsc{b} < \tau_\tsc{b}^*, \\[2mm]
    \proj{a_\tsc{a}}{a_\tsc{a}}\oplus \tfrac{1}{2}\mathds{1}_\tsc{b}\oplus (\proj{a_\tsc{a}}{a_\tsc{a}}\otimes\proj{-a_\tsc{b}}{-a_\tsc{b}}) & \text{if   } \tau_\tsc{a} \geq \tau_\tsc{a}^*, \tau_\tsc{b} < \tau_\tsc{b}^*, \\[2mm]
    \tfrac{1}{2}\mathds{1}_\tsc{a} \oplus \proj{b_\tsc{b}}{b_\tsc{b}}\oplus (\proj{-b_\tsc{a}}{-b_\tsc{a}}\otimes\proj{b_\tsc{b}}{b_\tsc{b}}) & \text{if   } \tau_\tsc{a} < \tau_\tsc{a}^*, \tau_\tsc{b} \geq \tau_\tsc{b}^*, \\[2mm]
    \proj{a_\tsc{a}}{a_\tsc{a}}\oplus \proj{b_\tsc{b}}{b_\tsc{b}}\oplus (\proj{a_\tsc{a}}{a_\tsc{a}}\otimes\proj{b_\tsc{b}}{b_\tsc{b}}) & \text{if   } \tau_\tsc{a} \geq \tau_\tsc{a}^*, \tau_\tsc{b} \geq \tau_\tsc{b}^*.
    \end{array} \right.
\end{equation}\\
Eq.~\eqref{Eq: time-dependent polystate EPR} prescribes a covariant sequence of updates that respects causality. The average of the product of outcomes can be computed as follows: let $\tau_\tsc{a} < \tau_\tsc{a}^*$ and $\tau_\tsc{b} < \tau_\tsc{b}^*$,
\begin{align}
    & \langle ab \rangle = \Tr[(\hat\sigma_{z,\tsc{a}} \otimes \hat\sigma_{\bm n, \tsc{b}}) \,\tilde\rho_{ab}(\tau_\tsc{a},\tau_\tsc{b})] = \bra{\Psi^-} \hat\sigma_{z,\tsc{a}} \otimes \hat\sigma_{\bm n, \tsc{b}} \ket{\Psi^-} = -\cos\theta_\tsc{ab},
\end{align}
which recovers the standard prediction of quantum mechanics for a Bell test. This is an \textit{a priori} expectation value, i.e., computed before the experiment is performed, which is why we need to use $\tau_\tsc{a} < \tau_\tsc{a}^*$ and $\tau_\tsc{b} < \tau_\tsc{b}^*$. More in general, the joint \textit{a priori} probability distribution is given by
\begin{align}
    & \text{Prob}(a,b) = \Tr[(\proj{a_\tsc{a}}{a_\tsc{a}} \otimes \proj{b_\tsc{b}}{b_\tsc{b}}) \,\tilde\rho_{ab}(\tau_\tsc{a},\tau_\tsc{b})] = \frac{1}{2}\Big[\delta_{ab}\sin^2\theta_\tsc{ab}/2  + (1-\delta_{ab})\cos^2\theta_\tsc{ab}/2 \Big].
\end{align}
The corresponding marginal probabilities can be computed as
\begin{align}
    \text{Prob}(a) & = \Tr[\proj{a_\tsc{a}}{a_\tsc{a}} \,\tilde\rho_{ab}(\tau_\tsc{a},\tau_\tsc{b}')] = \bra{a_\tsc{a}} \tfrac{1}{2}\mathds{1}_\tsc{a} \ket{a_\tsc{a}} = \tfrac{1}{2}, \\
    \text{Prob}(b) & = \Tr[\proj{b_\tsc{b}}{b_\tsc{b}} \,\tilde\rho_{ab}(\tau_\tsc{a}',\tau_\tsc{b})] = \bra{b_\tsc{b}} \tfrac{1}{2}\mathds{1}_\tsc{b} \ket{b_\tsc{b}} = \tfrac{1}{2},
\end{align}
where now $\tau_\tsc{b}'$ and $\tau_\tsc{a}'$ are arbitrary real numbers, since the expectation values of individual observables of A and B do not depend on the proper times we choose for B and A, respectively. We can also compute the marginal probabilities \textit{conditioned} on the outcome of the other experimenter: this implies computing the probability of a local experiment but taking into account non-local information from the other experimenter's outcome. Namely,
\begin{align}
    \text{Prob}(a | b) & = \Tr[(\proj{a_\tsc{a}}{a_\tsc{a}} \otimes \mathds{1}_\tsc{b}) \,\tilde\rho_{ab}(\tau_\tsc{a},\tau_\tsc{b}^*)] = |\!\braket{-b_\tsc{a}}{a_\tsc{a}}\!|^2 = \delta_{ab}\sin^2\theta_\tsc{ab}/2  + (1-\delta_{ab})\cos^2\theta_\tsc{ab}/2, \\
    \text{Prob}(b | a) & = \Tr[(\mathds{1}_\tsc{a} \otimes \proj{b_\tsc{b}}{b_\tsc{b}}) \,\tilde\rho_{ab}(\tau_\tsc{a}^*,\tau_\tsc{b})] = |\!\braket{-a_\tsc{b}}{b_\tsc{b}}\!|^2 = \delta_{ab}\sin^2\theta_\tsc{ab}/2  + (1-\delta_{ab})\cos^2\theta_\tsc{ab}/2.
\end{align}
This illustrates the difference, already mentioned in the main text, between individual operators of the form $\hat A \in \mathcal L(\hilb_\tsc{a})$ and joint operators of the form $\hat A \otimes \mathds{1}_\tsc{b} \in \mathcal L(\hilb_\tsc{a}\otimes \hilb_\tsc{b})$ (and analogously for B) in the polyperspective formalism: while the former provide marginal distributions, the latter provide conditional distributions.

\section{EPR pairs from a Cauchy surface perspective}~\label{Appendix: Cauchy}

\begin{figure}[h!]
\begin{center}
\includegraphics[width = 7cm]{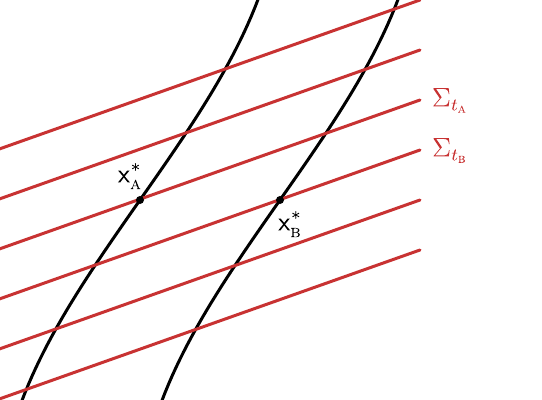}
\includegraphics[width = 7cm]{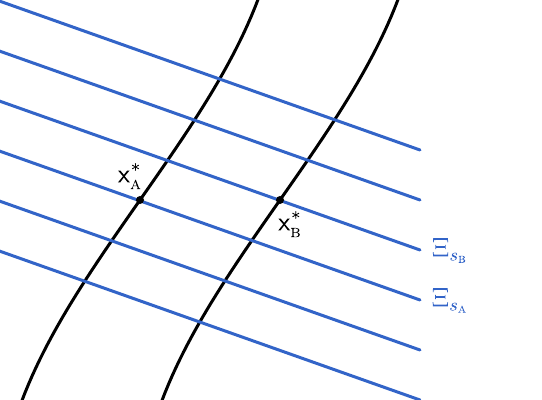}
\caption{A scenario where Alice and Bob perform spacelike measurements at $\msf x_\tc{a}^*$ and $\msf x_\tc{b}^*$, depicting two foliations $\Sigma_t$ and $\Xi_s$, where Alice's measurement happens ``before'' Bob's, according to the time parameter $s$, but ``after'' Bob's according to the time parameter $t$.}
\label{Fig: foliations}
\end{center}
\end{figure}

To illustrate the claim after Eq.~\eqref{Eq: foliation-dependent state} in the main text, let us consider the setup where Alice and Bob share a Bell pair and perform different operations at spacelike-separated regions from the perspective of foliation-dependent states. Consider that Alice measures $\hat\sigma_z$ at $\tau_\tc{a} = \tau_\tc{a}^*$, and Bob measures $\hat\sigma_x$ at $\tau_\tc{b} = \tau_\tc{b}^*$, such that $\msf x_\tc{a}^* \equiv \msf x_\tc{a}(\tau_\tc{a}^*)$ is spacelike separated from $\msf x_\tc{b}^* \equiv \msf x_\tc{b}(\tau_\tc{b}^*)$. In this case there is no well-defined causal order between $\msf x_\tc{a}^*$ and $\msf x_\tc{b}^*$. Specifically, there exist two foliations, $\{\Sigma_t\}$ and $\{\Xi_s\}$, where $t$ and $s$ are timelike parameters that parametrize each foliation, and let $t_\tc{a}$, $t_\tc{b}$, $s_\tc{a}$ and $s_\tc{b}$ be such that $\msf x_\tc{a}^*\in \Sigma_{t_\tc{a}}$, $\msf x_\tc{b}^*\in \Sigma_{t_\tc{b}}$, $\msf x_\tc{a}^*\in \Xi_{s_\tc{a}}$, $\msf x_\tc{b}^*\in \Xi_{s_\tc{b}}$. Given that $\msf x_\tc{a}^*$ and $\msf x_\tc{b}^*$ are spacelike separated, we can consider a scenario where $s_\tc{a}<s_\tc{b}$, while $t_\tc{a}>t_\tc{b}$ (see Fig.~\ref{Fig: foliations}).

Given Alice and Bob's measurement outcomes, we can now compute the foliation-dependent states $\hat{\rho}_\Sigma(t)$ and $\hat{\rho}_{\Xi}(s)$ for each value of $t$ and $s$, and compare them to the polystates they stem from. The initial polystate that describes the system before any operation is performed (i.e., for $\tau_\tsc{a} < \tau_\tsc{a}^*$ and $\tau_\tsc{b} < \tau_\tsc{b}^*$) is $\tilde{\rho}_0 =  \tfrac{1}{2}\mathds{1}_\tc{a}\oplus\tfrac{1}{2}\mathds{1}_\tc{b}\oplus\ket{\Psi^+}\!\!\bra{\Psi^+}$. For concreteness, we will consider the case when both Alice and Bob's outcomes are $+1$.

Let us first analyze the state described with respect to the foliation $\Sigma$, $\hat{\rho}_\Sigma(t)$. With respect to this foliation, the state is updated twice, first at $t_\tc{b}$, and then at $t_\tc{a}>t_\tc{b}$. When Bob performs his measurement at $t_\tc{b}$ and obtains $\ket{+_\tc{b}}$, the state at the surface is then updated to $\ket{+_\tc{a}+_\tc{b}}$, incorporating the total information available at $\Sigma_{t_\tc{b}}$. On the other hand, the polystate becomes $\tilde{\rho}(\tau_\tc{a}(t),\tau_\tc{b}(t)) = \tfrac{1}{2} \mathds{1}_\tc{a}\oplus \ket{+_\tc{b}}\!\!\bra{+_\tc{b}}\oplus \ket{+_\tc{a}+_\tc{b}}\!\!\bra{+_\tc{a}+_\tc{b}}$ for $t \in [t_\tsc{b}, t_\tsc{a})$. At $t_\tc{a}$, Alice measures $\hat\sigma_{z,\tsc{a}}$ and obtains $+1$. The state $\hat{\rho}_\Sigma$ is then updated to $\hat{\rho}_\Sigma(t_\tc{a}) = \ket{0_\tc{a}\!+_\tc{b}}\!\!\bra{0_\tc{a}\!+_\tc{b}}$, while the polystate becomes $\tilde{\rho}(\tau_\tc{a}(t),\tau_\tc{b}(t)) = \ket{0_\tc{a}}\!\!\bra{0_\tc{a}}\oplus\ket{+_\tc{b}}\!\!\bra{+_\tc{b}}\oplus \ket{0_\tc{a}\!+_\tc{b}}\!\!\bra{0_\tc{a}\!+_\tc{b}}$ for $t \geq t_\tsc{a}$. Notice that the fact that Alice obtained $+1$ is compatible with the statistics provided both by her individual state before the measurement and the polystate, as both $\hat{\rho}_\tc{a}(\tau_\tsc{a}(t)) = \tfrac{1}{2}\mathds{1}_\tc{a}$ and $\Tr_\tc{b}(\hat{\rho}_\Sigma(t)) = \ket{+_\tc{a}}\!\!\bra{+_\tc{a}}$ (for $t_\tc{b}\leq t <t_\tc{a}$) yield non-zero probabilities of obtaining $+1$ after measuring $\hat\sigma_z$. Overall, the states $\hat{\rho}_\Sigma(t)$ and $\tilde{\rho}(\tau_\tc{a}(t),\tau_\tc{b}(t))$ can be written as
\begin{equation}
    \hat{\rho}_\Sigma(t) = \begin{cases}
        \ket{\Psi^+}\!\!\bra{\Psi^+} &\text{if }t< t_\tc{b}\\[2mm]
        \ket{+_\tc{a}+_\tc{b}}\!\!\bra{+_\tc{a}+_\tc{b}} &\text{if }t_\tc{b}\leq t< t_\tc{a}\\[2mm]
        \ket{0_\tc{a}\!+_\tc{b}}\!\!\bra{0_\tc{a}+_\tc{b}} &\text{if }t \geq t_\tc{a}
    \end{cases}, \quad \tilde{\rho}(\tau_\tc{a}(t),\tau_\tc{b}(t)) = \begin{cases}
        \tfrac{1}{2}\mathds{1}_\tc{a}\oplus\tfrac{1}{2}\mathds{1}_\tc{b}\oplus\ket{\Psi^+}\!\!\bra{\Psi^+} &\text{if }t< t_\tc{b}\\[2mm]
        \tfrac{1}{2}\mathds{1}_\tc{a}\oplus \ket{+_\tc{b}}\!\!\bra{+_\tc{b}}\oplus\ket{+_\tc{a}+_\tc{b}}\!\!\bra{+_\tc{a}+_\tc{b}} &\text{if }t_\tc{b}\leq t <t_\tc{a}\\[2mm]
        \ket{0_\tc{a}}\!\!\bra{0_\tc{a}}\oplus \ket{+_\tc{b}}\!\!\bra{+_\tc{b}}\oplus\ket{0_\tc{a}\!+_\tc{b}}\!\!\bra{0_\tc{a}\!+_\tc{b}} &\text{if }t \geq t_\tc{a}
    \end{cases}.
\end{equation}

Let us now consider the state described with respect to the foliation $\Xi$, $\hat{\rho}_\Xi(s)$. With respect to this foliation, the state is first updated at $s_\tc{a}$, and then at $s_\tc{b}>s_\tc{a}$. Alice's measurement takes place at $s_\tc{a}$, and after obtaining an outcome of $+1$, the joint state is updated to $\ket{0_\tc{a}1_\tc{b}}$, incorporating the information available at $\Xi_{s_\tc{a}}$. The corresponding polystate $\tilde{\rho}(\tau_\tc{a}(s),\tau_\tc{b}(s))$ becomes $\tilde{\rho}(\tau_\tc{a}(s),\tau_\tc{b}(s)) = \ket{0_\tc{a}}\!\!\bra{0_\tc{a}}\oplus\tfrac{1}{2} \mathds{1}_\tc{b} \oplus\ket{0_\tc{a}1_\tc{b}}\!\!\bra{0_\tc{a}1_\tc{b}}$ for $s \in [s_\tsc{a},s_\tsc{b})$. At $s_\tc{b}$, Bob measures $\hat\sigma_{x,\tsc{b}}$ and obtains $+1$, updating the state $\hat{\rho}_\Xi$ to $\hat{\rho}_\Xi(s_\tc{b}) = \ket{0_\tc{a}\!+_\tc{b}}\!\!\bra{0_\tc{a}\!+_\tc{b}}$, with $\tilde{\rho}(\tau_\tc{a}(s),\tau_\tc{b}(s)) = \ket{0_\tc{a}}\!\!\bra{0_\tc{a}}\oplus\ket{+_\tc{b}}\!\!\bra{+_\tc{b}}\oplus \ket{0_\tc{a}\!+_\tc{b}}\!\!\bra{0_\tc{a}\!+_\tc{b}}$ for $s \geq s_\tsc{b}$. In this case we also have that the state obtained after Bob's measurement is compatible with the statistics provided both by his individual state before the measurement and by the polystate, given that for $s_\tc{a}\leq s <s_\tc{b}$, $\hat{\rho}_\tc{b}(\tau_\tsc{b}(s)) = \tfrac{1}{2}\mathds{1}_\tc{b}$ and $\Tr_\tc{a}(\hat{\rho}_\Xi(t)) = \ket{1_\tc{b}}\!\!\bra{1_\tc{b}}$, which both yield non-zero probabilities of obtaining $+1$ after measuring $\hat\sigma_x$. Altogether, the states $\hat{\rho}_\Xi(s)$ and $\tilde{\rho}(\tau_\tc{a}(s),\tau_\tc{b}(s))$ can be written as
\begin{equation}
    \hat{\rho}_\Xi(s) = \begin{cases}
        \ket{\Psi^+}\!\!\bra{\Psi^+} &\text{if }s< s_\tc{a}\\[2mm]
        \ket{0_\tc{a}1_\tc{b}}\!\!\bra{0_\tc{a}1_\tc{b}} &\text{if }s_\tc{a}\leq s< s_\tc{b}\\[2mm]
        \ket{0_\tc{a}+_\tc{b}}\!\!\bra{0_\tc{a}+_\tc{b}} &\text{if }s \geq s_\tc{b}
    \end{cases}, \quad \tilde{\rho}(\tau_\tc{a}(s),\tau_\tc{b}(s)) = \begin{cases}
        \tfrac{1}{2}\mathds{1}_\tc{a}\oplus\tfrac{1}{2}\mathds{1}_\tc{b}\oplus\ket{\Psi^+}\!\!\bra{\Psi^+} &\text{if }s<s_\tc{a}\\[2mm]
        \ket{0_\tc{a}}\!\!\bra{0_\tc{a}}\oplus\tfrac{1}{2}\mathds{1}_\tc{b}\oplus\ket{0_\tc{a}1_\tc{b}}\!\!\bra{0_\tc{a}1_\tc{b}} &\text{if }s_\tc{a}\leq s <s_\tc{b}\\[2mm]
        \ket{0_\tc{a}}\!\!\bra{0_\tc{a}}\oplus \ket{+_\tc{b}}\!\!\bra{+_\tc{b}}\oplus\ket{0_\tc{a}\!+_\tc{b}}\!\!\bra{0_\tc{a}\!+_\tc{b}} &\text{if } s \geq s_\tc{b}
    \end{cases}.
\end{equation}
Notice that after both measurements are performed, the states $\hat{\rho}_\Sigma(t)$ and $\hat{\rho}_\Xi(s)$ agree whenever $t\geq t_\tc{a}$ and $s\geq s_\tc{b}$. This is because, after these times, all operations performed on the qubits lie in the causal past of both $\Sigma_t$ and $\Xi_s$, so that the information about their outcomes is encoded in both states. Equivalently, this agreement can be understood as due to the fact that for $t\geq t_\tc{a}$ and $s\geq s_\tc{b}$, we also have $\tau_\tsc{a} \geq \tau_\tsc{a}^*$ and $\tau_\tsc{b} \geq \tau_\tsc{b}^*$. This implies that \mbox{$\{\msf x_\tsc{a}^*\} \cup \{\msf x_\tsc{b}^*\} \subset \mathcal J^-(\msf x_\tsc{a}(\tau_\tsc{a})) \cup \mathcal J^-(\msf x_\tsc{b}(\tau_\tsc{b}))$}, and therefore the polystate is  \mbox{$\tilde\rho(\tau_\tsc{a},\tau_\tsc{b}) = \ket{0_\tc{a}}\!\!\bra{0_\tc{a}}\oplus \ket{+_\tc{b}}\!\!\bra{+_\tc{b}}\oplus\ket{0_\tc{a}\!+_\tc{b}}\!\!\bra{0_\tc{a}\!+_\tc{b}}$}.

\twocolumngrid

\bibliography{references}

\begin{thebibliography}{27}%
\makeatletter
\providecommand \@ifxundefined [1]{%
 \@ifx{#1\undefined}
}%
\providecommand \@ifnum [1]{%
 \ifnum #1\expandafter \@firstoftwo
 \else \expandafter \@secondoftwo
 \fi
}%
\providecommand \@ifx [1]{%
 \ifx #1\expandafter \@firstoftwo
 \else \expandafter \@secondoftwo
 \fi
}%
\providecommand \natexlab [1]{#1}%
\providecommand \enquote  [1]{``#1''}%
\providecommand \bibnamefont  [1]{#1}%
\providecommand \bibfnamefont [1]{#1}%
\providecommand \citenamefont [1]{#1}%
\providecommand \href@noop [0]{\@secondoftwo}%
\providecommand \href [0]{\begingroup \@sanitize@url \@href}%
\providecommand \@href[1]{\@@startlink{#1}\@@href}%
\providecommand \@@href[1]{\endgroup#1\@@endlink}%
\providecommand \@sanitize@url [0]{\catcode `\\12\catcode `\$12\catcode `\&12\catcode `\#12\catcode `\^12\catcode `\_12\catcode `\%12\relax}%
\providecommand \@@startlink[1]{}%
\providecommand \@@endlink[0]{}%
\providecommand \url  [0]{\begingroup\@sanitize@url \@url }%
\providecommand \@url [1]{\endgroup\@href {#1}{\urlprefix }}%
\providecommand \urlprefix  [0]{URL }%
\providecommand \Eprint [0]{\href }%
\providecommand \doibase [0]{https://doi.org/}%
\providecommand \selectlanguage [0]{\@gobble}%
\providecommand \bibinfo  [0]{\@secondoftwo}%
\providecommand \bibfield  [0]{\@secondoftwo}%
\providecommand \translation [1]{[#1]}%
\providecommand \BibitemOpen [0]{}%
\providecommand \bibitemStop [0]{}%
\providecommand \bibitemNoStop [0]{.\EOS\space}%
\providecommand \EOS [0]{\spacefactor3000\relax}%
\providecommand \BibitemShut  [1]{\csname bibitem#1\endcsname}%
\let\auto@bib@innerbib\@empty
\bibitem [{\citenamefont {L\"{u}ders}(1951)}]{Luders1951}%
  \BibitemOpen
  \bibfield  {author} {\bibinfo {author} {\bibfnamefont {G.}~\bibnamefont {L\"{u}ders}},\ }\bibfield  {title} {\bibinfo {title} {Concerning the state-change due to the measurement process},\ }\href {https://doi.org/10.1002/andp.200610207} {\bibfield  {journal} {\bibinfo  {journal} {Ann. Phys. (Berl.)}\ }\textbf {\bibinfo {volume} {15}},\ \bibinfo {pages} {663} (\bibinfo {year} {1951})}\BibitemShut {NoStop}%
\bibitem [{\citenamefont {Jubb}(2022)}]{Jubb2022}%
  \BibitemOpen
  \bibfield  {author} {\bibinfo {author} {\bibfnamefont {I.}~\bibnamefont {Jubb}},\ }\bibfield  {title} {\bibinfo {title} {Causal state updates in real scalar quantum field theory},\ }\href {https://doi.org/10.1103/PhysRevD.105.025003} {\bibfield  {journal} {\bibinfo  {journal} {Phys. Rev. D}\ }\textbf {\bibinfo {volume} {105}},\ \bibinfo {pages} {025003} (\bibinfo {year} {2022})}\BibitemShut {NoStop}%
\bibitem [{\citenamefont {Albertini}\ and\ \citenamefont {Jubb}(2023)}]{Albertini2023}%
  \BibitemOpen
  \bibfield  {author} {\bibinfo {author} {\bibfnamefont {E.}~\bibnamefont {Albertini}}\ and\ \bibinfo {author} {\bibfnamefont {I.}~\bibnamefont {Jubb}},\ }\href@noop {} {\bibinfo {title} {{Are Ideal Measurements of Real Scalar Fields Causal?}}} (\bibinfo {year} {2023}),\ \Eprint {https://arxiv.org/abs/2306.12980} {arXiv:2306.12980 [quant-ph]} \BibitemShut {NoStop}%
\bibitem [{\citenamefont {Fewster}\ and\ \citenamefont {Verch}(2020)}]{FewsterVerch}%
  \BibitemOpen
  \bibfield  {author} {\bibinfo {author} {\bibfnamefont {C.~J.}\ \bibnamefont {Fewster}}\ and\ \bibinfo {author} {\bibfnamefont {R.}~\bibnamefont {Verch}},\ }\bibfield  {title} {\bibinfo {title} {Quantum {F}ields and {L}ocal {M}easurements},\ }\href {https://doi.org/10.1007/s00220-020-03800-6} {\bibfield  {journal} {\bibinfo  {journal} {Commun. Math. Phys.}\ }\textbf {\bibinfo {volume} {378}},\ \bibinfo {pages} {851} (\bibinfo {year} {2020})}\BibitemShut {NoStop}%
\bibitem [{\citenamefont {Fewster}(2020)}]{Fewster2020covariant}%
  \BibitemOpen
  \bibfield  {author} {\bibinfo {author} {\bibfnamefont {C.~J.}\ \bibnamefont {Fewster}},\ }\bibinfo {title} {{A Generally Covariant Measurement Scheme for Quantum Field Theory in Curved Spacetimes}},\ in\ \href {https://doi.org/10.1007/978-3-030-38941-3_11} {\emph {\bibinfo {booktitle} {Progress and Visions in Quantum Theory in View of Gravity}}}\ (\bibinfo  {publisher} {Springer International Publishing},\ \bibinfo {year} {2020})\ pp.\ \bibinfo {pages} {253--268}\BibitemShut {NoStop}%
\bibitem [{\citenamefont {Smith}(2019)}]{AlexSmith2019Thesis}%
  \BibitemOpen
  \bibfield  {author} {\bibinfo {author} {\bibfnamefont {A.~R.~H.}\ \bibnamefont {Smith}},\ }\href {https://doi.org/10.1007/978-3-030-11000-0} {\emph {\bibinfo {title} {{Detectors, Reference Frames, and Time}}}}\ (\bibinfo  {publisher} {Springer International Publishing},\ \bibinfo {year} {2019})\BibitemShut {NoStop}%
\bibitem [{\citenamefont {Polo-G\'omez}\ \emph {et~al.}(2022)\citenamefont {Polo-G\'omez}, \citenamefont {Garay},\ and\ \citenamefont {Mart\'{\i}n-Mart\'{\i}nez}}]{MeasurementTheory}%
  \BibitemOpen
  \bibfield  {author} {\bibinfo {author} {\bibfnamefont {J.}~\bibnamefont {Polo-G\'omez}}, \bibinfo {author} {\bibfnamefont {L.~J.}\ \bibnamefont {Garay}},\ and\ \bibinfo {author} {\bibfnamefont {E.}~\bibnamefont {Mart\'{\i}n-Mart\'{\i}nez}},\ }\bibfield  {title} {\bibinfo {title} {A detector-based measurement theory for quantum field theory},\ }\href {https://doi.org/10.1103/PhysRevD.105.065003} {\bibfield  {journal} {\bibinfo  {journal} {Phys. Rev. D}\ }\textbf {\bibinfo {volume} {105}},\ \bibinfo {pages} {065003} (\bibinfo {year} {2022})}\BibitemShut {NoStop}%
\bibitem [{\citenamefont {Maeso-Garc\'{\i}a}\ \emph {et~al.}(2023)\citenamefont {Maeso-Garc\'{\i}a}, \citenamefont {Polo-G\'omez},\ and\ \citenamefont {Mart\'{\i}n-Mart\'{\i}nez}}]{MeasurementsEH}%
  \BibitemOpen
  \bibfield  {author} {\bibinfo {author} {\bibfnamefont {H.}~\bibnamefont {Maeso-Garc\'{\i}a}}, \bibinfo {author} {\bibfnamefont {J.}~\bibnamefont {Polo-G\'omez}},\ and\ \bibinfo {author} {\bibfnamefont {E.}~\bibnamefont {Mart\'{\i}n-Mart\'{\i}nez}},\ }\bibfield  {title} {\bibinfo {title} {How measuring a quantum field affects entanglement harvesting},\ }\href {https://doi.org/10.1103/PhysRevD.107.045011} {\bibfield  {journal} {\bibinfo  {journal} {Phys. Rev. D}\ }\textbf {\bibinfo {volume} {107}},\ \bibinfo {pages} {045011} (\bibinfo {year} {2023})}\BibitemShut {NoStop}%
\bibitem [{\citenamefont {Grimmer}\ \emph {et~al.}(2023)\citenamefont {Grimmer}, \citenamefont {Melgarejo-Lermas}, \citenamefont {Polo-G\'omez},\ and\ \citenamefont {Mart\'in-Mart\'inez}}]{DanIreneML}%
  \BibitemOpen
  \bibfield  {author} {\bibinfo {author} {\bibfnamefont {D.}~\bibnamefont {Grimmer}}, \bibinfo {author} {\bibfnamefont {I.}~\bibnamefont {Melgarejo-Lermas}}, \bibinfo {author} {\bibfnamefont {J.}~\bibnamefont {Polo-G\'omez}},\ and\ \bibinfo {author} {\bibfnamefont {E.}~\bibnamefont {Mart\'in-Mart\'inez}},\ }\bibfield  {title} {\bibinfo {title} {Decoding quantum field theory with machine learning},\ }\href {https://doi.org/10.1007/JHEP08(2023)031} {\bibfield  {journal} {\bibinfo  {journal} {J. High Energy Phys.}\ }\textbf {\bibinfo {volume} {2023}}\bibinfo  {number} { (8)},\ \bibinfo {pages} {31}}\BibitemShut {NoStop}%
\bibitem [{\citenamefont {Oeckl}(2013)}]{Oeckl2013}%
  \BibitemOpen
\bibfield  {number} {  }\bibfield  {author} {\bibinfo {author} {\bibfnamefont {R.}~\bibnamefont {Oeckl}},\ }\bibfield  {title} {\bibinfo {title} {{A Positive Formalism for Quantum Theory in the General Boundary Formulation}},\ }\href {https://doi.org/10.1007/s10701-013-9741-5} {\bibfield  {journal} {\bibinfo  {journal} {Found. Phys.}\ }\textbf {\bibinfo {volume} {43}},\ \bibinfo {pages} {1206} (\bibinfo {year} {2013})}\BibitemShut {NoStop}%
\bibitem [{\citenamefont {Oeckl}(2019)}]{Oeckl2019}%
  \BibitemOpen
  \bibfield  {author} {\bibinfo {author} {\bibfnamefont {R.}~\bibnamefont {Oeckl}},\ }\bibfield  {title} {\bibinfo {title} {{A local and operational framework for the foundations of physics}},\ }\href {https://doi.org/10.4310/atmp.2019.v23.n2.a4} {\bibfield  {journal} {\bibinfo  {journal} {Adv. Theor. Math. Phys.}\ }\textbf {\bibinfo {volume} {23}},\ \bibinfo {pages} {437} (\bibinfo {year} {2019})}\BibitemShut {NoStop}%
\bibitem [{\citenamefont {Oeckl}(2025)}]{Oeckl2025Spectral}%
  \BibitemOpen
  \bibfield  {author} {\bibinfo {author} {\bibfnamefont {R.}~\bibnamefont {Oeckl}},\ }\bibfield  {title} {\bibinfo {title} {{Spectral decomposition of field operators and causal measurement in quantum field theory}},\ }\href {https://doi.org/10.1063/5.0245368} {\bibfield  {journal} {\bibinfo  {journal} {J. Math. Phys.}\ }\textbf {\bibinfo {volume} {66}},\ \bibinfo {pages} {042302} (\bibinfo {year} {2025})}\BibitemShut {NoStop}%
\bibitem [{\citenamefont {Oeckl}\ and\ \citenamefont {Zampeli}(2025)}]{OeacklZampeli2025}%
  \BibitemOpen
  \bibfield  {author} {\bibinfo {author} {\bibfnamefont {R.}~\bibnamefont {Oeckl}}\ and\ \bibinfo {author} {\bibfnamefont {A.}~\bibnamefont {Zampeli}},\ }\href@noop {} {\bibinfo {title} {{Towards local and compositional measurements in quantum field theory}}} (\bibinfo {year} {2025}),\ \Eprint {https://arxiv.org/abs/2505.10968} {arXiv:2505.10968 [hep-th]} \BibitemShut {NoStop}%
\bibitem [{\citenamefont {Sorkin}(1993)}]{Sorkin1993}%
  \BibitemOpen
  \bibfield  {author} {\bibinfo {author} {\bibfnamefont {R.~D.}\ \bibnamefont {Sorkin}},\ }\href@noop {} {\bibinfo {title} {Impossible {M}easurements on {Q}uantum {F}ields}} (\bibinfo {year} {1993}),\ \Eprint {https://arxiv.org/abs/gr-qc/9302018} {arXiv:gr-qc/9302018} \BibitemShut {NoStop}%
\bibitem [{\citenamefont {Aharonov}\ and\ \citenamefont {Albert}(1981)}]{AharonovAlbert1981}%
  \BibitemOpen
  \bibfield  {author} {\bibinfo {author} {\bibfnamefont {Y.}~\bibnamefont {Aharonov}}\ and\ \bibinfo {author} {\bibfnamefont {D.~Z.}\ \bibnamefont {Albert}},\ }\bibfield  {title} {\bibinfo {title} {{Can we make sense out of the measurement process in relativistic quantum mechanics?}},\ }\href {https://doi.org/10.1103/PhysRevD.24.359} {\bibfield  {journal} {\bibinfo  {journal} {Phys. Rev. D}\ }\textbf {\bibinfo {volume} {24}},\ \bibinfo {pages} {359} (\bibinfo {year} {1981})}\BibitemShut {NoStop}%
\bibitem [{\citenamefont {Hellwig}\ and\ \citenamefont {Kraus}(1970)}]{Hellwig1970formal}%
  \BibitemOpen
  \bibfield  {author} {\bibinfo {author} {\bibfnamefont {K.~E.}\ \bibnamefont {Hellwig}}\ and\ \bibinfo {author} {\bibfnamefont {K.}~\bibnamefont {Kraus}},\ }\bibfield  {title} {\bibinfo {title} {Formal {D}escription of {M}easurements in {L}ocal {Q}uantum {F}ield {T}heory},\ }\href {https://doi.org/10.1103/PhysRevD.1.566} {\bibfield  {journal} {\bibinfo  {journal} {Phys. Rev. D}\ }\textbf {\bibinfo {volume} {1}},\ \bibinfo {pages} {566} (\bibinfo {year} {1970})}\BibitemShut {NoStop}%
\bibitem [{\citenamefont {Aharonov}\ and\ \citenamefont {Albert}(1980)}]{AharonovAlbert1980}%
  \BibitemOpen
  \bibfield  {author} {\bibinfo {author} {\bibfnamefont {Y.}~\bibnamefont {Aharonov}}\ and\ \bibinfo {author} {\bibfnamefont {D.~Z.}\ \bibnamefont {Albert}},\ }\bibfield  {title} {\bibinfo {title} {{States and observables in relativistic quantum field theories}},\ }\href {https://doi.org/10.1103/PhysRevD.21.3316} {\bibfield  {journal} {\bibinfo  {journal} {Phys. Rev. D}\ }\textbf {\bibinfo {volume} {21}},\ \bibinfo {pages} {3316} (\bibinfo {year} {1980})}\BibitemShut {NoStop}%
\bibitem [{\citenamefont {Fewster}\ and\ \citenamefont {Verch}(2025)}]{Fewster2025MeasurementinQFTEncyclopedia}%
  \BibitemOpen
  \bibfield  {author} {\bibinfo {author} {\bibfnamefont {C.~J.}\ \bibnamefont {Fewster}}\ and\ \bibinfo {author} {\bibfnamefont {R.}~\bibnamefont {Verch}},\ }\bibinfo {title} {{Measurement in Quantum Field Theory}},\ in\ \href {https://doi.org/10.1016/b978-0-323-95703-8.00076-8} {\emph {\bibinfo {booktitle} {Encyclopedia of Mathematical Physics}}}\ (\bibinfo  {publisher} {Elsevier},\ \bibinfo {year} {2025})\ pp.\ \bibinfo {pages} {335--345}\BibitemShut {NoStop}%
\bibitem [{\citenamefont {Wald}(1984)}]{Wald1984GR}%
  \BibitemOpen
  \bibfield  {author} {\bibinfo {author} {\bibfnamefont {R.~M.}\ \bibnamefont {Wald}},\ }\href {https://doi.org/10.7208/chicago/9780226870373.001.0001} {\emph {\bibinfo {title} {{General Relativity}}}}\ (\bibinfo  {publisher} {Chicago University Press.},\ \bibinfo {year} {1984})\BibitemShut {NoStop}%
\bibitem [{\citenamefont {Pranzini}\ and\ \citenamefont {Keski-Vakkuri}(2025)}]{Pranzini2023}%
  \BibitemOpen
  \bibfield  {author} {\bibinfo {author} {\bibfnamefont {N.}~\bibnamefont {Pranzini}}\ and\ \bibinfo {author} {\bibfnamefont {E.}~\bibnamefont {Keski-Vakkuri}},\ }\bibfield  {title} {\bibinfo {title} {{Detector-based measurements for QFT: Two issues and an algebraic QFT proposal}},\ }\href {https://doi.org/10.1103/PhysRevD.111.045016} {\bibfield  {journal} {\bibinfo  {journal} {Phys. Rev. D}\ }\textbf {\bibinfo {volume} {111}},\ \bibinfo {pages} {045016} (\bibinfo {year} {2025})}\BibitemShut {NoStop}%
\bibitem [{\citenamefont {Fewster}(2025)}]{Fewster2025lectures}%
  \BibitemOpen
  \bibfield  {author} {\bibinfo {author} {\bibfnamefont {C.~J.}\ \bibnamefont {Fewster}},\ }\href@noop {} {\bibinfo {title} {{Lectures on measurement in quantum field theory}}} (\bibinfo {year} {2025}),\ \Eprint {https://arxiv.org/abs/2504.17437} {arXiv:2504.17437 [gr-qc]} \BibitemShut {NoStop}%
\bibitem [{\citenamefont {Ruep}(2021)}]{Ruep2021}%
  \BibitemOpen
  \bibfield  {author} {\bibinfo {author} {\bibfnamefont {M.~H.}\ \bibnamefont {Ruep}},\ }\bibfield  {title} {\bibinfo {title} {Weakly coupled local particle detectors cannot harvest entanglement},\ }\href {https://doi.org/10.1088/1361-6382/ac1b08} {\bibfield  {journal} {\bibinfo  {journal} {Class. Quantum Gravity}\ }\textbf {\bibinfo {volume} {38}},\ \bibinfo {pages} {195029} (\bibinfo {year} {2021})}\BibitemShut {NoStop}%
\bibitem [{\citenamefont {Ballentine}(1970)}]{Ballentine1970}%
  \BibitemOpen
  \bibfield  {author} {\bibinfo {author} {\bibfnamefont {L.~E.}\ \bibnamefont {Ballentine}},\ }\bibfield  {title} {\bibinfo {title} {The statistical interpretation of quantum mechanics},\ }\href {https://doi.org/10.1103/RevModPhys.42.358} {\bibfield  {journal} {\bibinfo  {journal} {Rev. Mod. Phys.}\ }\textbf {\bibinfo {volume} {42}},\ \bibinfo {pages} {358} (\bibinfo {year} {1970})}\BibitemShut {NoStop}%
\bibitem [{\citenamefont {Polo-G\'omez}\ \emph {et~al.}()\citenamefont {Polo-G\'omez}, \citenamefont {Garay},\ and\ \citenamefont {Mart\'in-Mart\'inez}}]{MeasurementTheoryUpdated}%
  \BibitemOpen
  \bibfield  {author} {\bibinfo {author} {\bibfnamefont {J.}~\bibnamefont {Polo-G\'omez}}, \bibinfo {author} {\bibfnamefont {L.~J.}\ \bibnamefont {Garay}},\ and\ \bibinfo {author} {\bibfnamefont {E.}~\bibnamefont {Mart\'in-Mart\'inez}},\ }\href@noop {} {\bibinfo {title} {in preparation}}\BibitemShut {NoStop}%
\bibitem [{\citenamefont {Malin}(1982)}]{Malin1982}%
  \BibitemOpen
  \bibfield  {author} {\bibinfo {author} {\bibfnamefont {S.}~\bibnamefont {Malin}},\ }\bibfield  {title} {\bibinfo {title} {Observer dependence of quantum states in relativistic quantum field theories},\ }\href {https://doi.org/10.1103/PhysRevD.26.1330} {\bibfield  {journal} {\bibinfo  {journal} {Phys. Rev. D}\ }\textbf {\bibinfo {volume} {26}},\ \bibinfo {pages} {1330} (\bibinfo {year} {1982})}\BibitemShut {NoStop}%
\bibitem [{\citenamefont {G{\l}owacki}(2024)}]{Glowacki2024}%
  \BibitemOpen
  \bibfield  {author} {\bibinfo {author} {\bibfnamefont {J.}~\bibnamefont {G{\l}owacki}},\ }\href@noop {} {\bibinfo {title} {{Towards Relational Quantum Field Theory}}} (\bibinfo {year} {2024}),\ \Eprint {https://arxiv.org/abs/2405.15455} {arXiv:2405.15455 [quant-ph]} \BibitemShut {NoStop}%
\bibitem [{\citenamefont {Dowker}(2011)}]{Dowker2011Useless}%
  \BibitemOpen
  \bibfield  {author} {\bibinfo {author} {\bibfnamefont {F.}~\bibnamefont {Dowker}},\ }\href@noop {} {\bibinfo {title} {{Useless Qubits in ``Relativistic Quantum Information''}}} (\bibinfo {year} {2011}),\ \Eprint {https://arxiv.org/abs/1111.2308} {arXiv:1111.2308 [quant-ph]} \BibitemShut {NoStop}%
\end{thebibliography}%

\end{document}